\documentclass[amsmath,amssymb,nofootinbib,showpacs,twocolumn,nofootinbib]{revtex4-1}
\usepackage{dcolumn}
\usepackage{bm}
\usepackage{graphicx}
\usepackage{epstopdf}
\usepackage{epsfig}
\usepackage{color}
\usepackage{wasysym}
\usepackage{natbib}
\usepackage{twoopt}
\usepackage{dashrule}
\definecolor{slateblue}{rgb}{0.2,0.2,0.6}

\usepackage[pdftex,breaklinks=true,colorlinks=true,linkcolor=slateblue,citecolor=slateblue,urlcolor=slateblue,pdfauthor={Tomassetti},pdftitle={PHevsTime}]{hyperref}
\def\MyTitle#1{\noindent{\textit{#1}}}
 
\definecolor{ColorTitle}{cmyk}{0,.88,.77,.40}

\newcommand{\ApJ}{ApJ}

\newcommand{\PRL}{Phys. Rev. Lett.}
\newcommand{\PRD}{Phys. Rev. D}

\newcommand{\etal}{et al.}
 
\newcommand{\ie}{\textit{i.e.}} 
\newcommand{\Paper}{Letter}

\newcommand{\R}{\ensuremath{\mathcal{R}}}
\newcommand{\p}{\textrm{p}}
\newcommand{\Hyd}{\textrm{H}}
\newcommand{\He}{\textrm{He}}

\newcommand{\Be}{\textrm{Be}}
\newcommand{\B}{\textrm{B}}
\newcommand{\C}{\textrm{C}}

\newcommand{\Oxy}{\textrm{O}}
\newcommand{\pHe}{\textrm{p/He}}
\newcommand{\BC}{\textrm{B}/\textrm{C}}
\begin{document}
\title{Testing diffusion of cosmic rays in the heliosphere\\ with proton and helium data from AMS}

\author{
N. Tomassetti$^{\,1}$, 
F. Bar\~{a}o$^{\,2}$, 
B. Bertucci$^{\,1}$, 
E. Fiandrini$^{\,1}$, 
J. L. Figueiredo$^{\,2}$,
J. B. Lousada$^{\,2}$,
M. Orcinha$^{\,2}$}

\address{$^{1}$\,Universit{\`a} degli Studi di Perugia \& INFN-Perugia, I-06100 Perugia, Italy;}
\address{$^{2}$\,Laborat{\'o}rio de Instrumenta\c{c}{\~a}o e F{\'i}sica Experimental de Part{\'i}culas, P-1000 Lisboa, Portugal}
\begin{abstract}
  After six years of continuous observations in space, the Alpha Magnetic Spectrometer experiment has released
  new data on the temporal evolution of the proton and helium fluxes in cosmic rays. 
  These data revealed that the ratio between proton and helium fluxes at the same value of rigidity $\R=p/Z$ (momentum/charge ratio)
  is not constant at $\R\lesssim$\,3\,GV. In particular, the ratio is found 
  to decrease steadily during the descending phase of Solar Cycle 24 toward the next minimum.
  We show that such a behavior is a remarkable signature of the $\beta\times\lambda(\R)$ dependence in the diffusion
  of cosmic rays in heliosphere, where $\beta$ is their adimensional speed and $\lambda(\R)$ is their mean free path,
  a \emph{universal} function of rigidity for all nuclei.
  This dependence is responsible for distinctive charge/mass dependent effects in the time-dependent modulation of low-rigidity particles.
\end{abstract}
\maketitle

%%%%%%%%%%%%%%%%%%%%%%%%%%%%%%
\MyTitle{Introduction} --- %%%
%%%%%%%%%%%%%%%%%%%%%%%%%%%%%%
%
Galactic cosmic rays (CRs) are high-energy charged particles that originate in violent astrophysical processes 
outside the solar system, such as supernova explosions or stellar winds.
When entering the heliosphere, CRs are deflected and decelerated by the turbulent magnetic fields of the Sun,
dragged out by the solar wind, that make their energy spectra significantly different from those in the interstellar space.
Moreover, the effect is not stationary, but it changes periodically with the Sun's 11-year activity cycle,
causing the anti-correlation between near-Earth CR fluxes and sunspot numbers.
The observed change of the CR flux over the solar activity cycle is known as \emph{solar modulation} of CRs in the heliosphere.
Along with its implications in solar or plasma astrophysics,
the solar modulation effect is an important factor for Galactic CR physics studies,
as it limits our ability to identify the CR sources, or to search for dark matter annihilation signals,
\ie, the main goals of the Alpha Magnetic Spectrometer (AMS) experiment \citep{Grenier2015}.
Besides, the varying CR flux in the interplanetary space provides a significant challenge for
space missions and air travelers \citep{Kudela2000,Tomassetti2017TimeLag}.

In the past few decades, a consistent paradigm of CR transport in the heliosphere has been established,
according to which the modulation effect
is caused by a combination of diffusion, drift, convection, and adiabatic cooling.
All these processes arise from the dynamics of particle gyromotion in large-scale magnetic fields
and scattering off its small-scale irregularities, and thus they are governed
by magnetic rigidity (momentum/charge ratio) $\R=p/Z$  \citep{Potgieter2013,Moraal2013}.
Observationally, valuable pieces of information have been gained 
thanks to the space missions CRIS/ACE \citep{Wiedenbeck2009}, IMP-7/8 \citep{GarciaMunoz1997},
\emph{Ulysses} \citep{Heber2009} and, more recently Voyager-1 \citep{Cummings2016}, EPHIN/SOHO \citep{Kuhl2016},
and PAMELA \citep{Adriani2013,Martucci2018,Bindi2017}.
Very recently, the Alpha Magnetic Spectrometer (AMS) experiment has released 
measurements of the temporal dependence of the proton and helium fluxes
in CRs between 2011 and 2017, unveiling new details of CR modulation \citep{Aguilar2018PHe}.
With an accuracy at the level of 1\,\% and a 27-day time resolution, AMS has observed temporal
variations in the CR fluxes up to $\R\sim$\,40\,GV. 
On monthly time scales, the measured proton and \He{} fluxes show nearly identical fine structures
in time and relative amplitude while, on yearly time scales,
little differences have emerged between the two species.
The \pHe{} ratio between proton and helium fluxes evaluated at the same
value of rigidity shows a long-term time dependence for $\R\lesssim$\,3\,GV.
In particular, a remarkable decrease of the \pHe{} ratio is observed between
March 2015 and May 2017, coinciding with the descending period of the Sun's
activity --toward the next solar minimum-- and the increase of both fluxes.
As noted by the AMS collaboration, explanations for this feature
may arise from mass/charge dependencies in CR transport at low-rigidity 
or from differences in the local interstellar spectra (LIS) of CR proton and helium,
with the possible influence of the $^{3}$\He--$^{4}$\He{} isotopes in the above
effects \citep{Aguilar2018PHe,Jokipii1967,Gloeckler1967,Biswas1967,Herbst2017,Gieseler2017}.

In this \Paper, we show that the long-term behavior of the \pHe{} ratio is a signature of
the $\beta\times\lambda(\R)$ dependence of CR diffusion in the interplanetary magnetic fields,
where $\beta$ is the adimensional speed of CRs and $\lambda$ is their mean free path. 
From the quasilinear theory of CR diffusion the mean free path of CR nuclei in the heliosphere
is connected to the spectrum of magnetic turbulence \citep{TeufelSchlickeiser2002},
and often described by a \emph{universal} function of rigidity $\lambda=\lambda(\R)$, \ie,
having the same $\R$-dependence for all CR nuclei with different mass or charge \citep{Moraal2013,Fisk1971}. 
The diffusion coefficient follows simply from $K=\beta\lambda/3$.
As we will show, this form causes observable differences in the time-dependent modulation
of low-$\R$ particles characterized by different $Z/A$ ratio between charge and mass number.
The nuclear ratio between CR protons and helium is the best suited to test this effect
because it maximizes the $Z/A$ difference between the two examined species and,
being a ratio of positively charged particles, it cancels out charge-sign dependent effects caused by drift.
Nonrelativistic proton ($Z/A=$1) and helium ($Z/A\approx$1/2) nuclei observed at the same
rigidity travel at different speed, with $\beta_{p}>\beta_{He}$.
Thus, cosmic protons must experience faster diffusion than heavier nuclei,
while in the high-$\R$ limit the diffusion of the two particles become identical.
To test the above suggestions against possible LIS-induced effects \citep{Gieseler2017},
a careful modeling at the precision level of the data is required.
In this work, to assess the \p-\He{} LIS's and their uncertainties, we make use of state-of-the-art
models of CR propagation in Galaxy constrained against new Voyager-1 data collected in the interstellar
space \citep{Cummings2016}, and high-energy measurements from AMS \citep{Aguilar2015Proton,Aguilar2015Helium}.
We also account for the isotopic composition of the hydrogen and helium fluxes,
including their uncertainties, and for the modulation of all relevant isotopes.
To compute the time-dependent effect of solar modulation, we make use of numerical
calculations of CR transport in the heliosphere calibrated against 
the new AMS data on monthly-resolved CR protons \citep{Aguilar2018PHe}.
This approach enables us to \emph{predict} the temporal dependence of the \pHe{} ratio
and to test specific forms of the diffusion coefficient.

%%%%%%%%%%%%%%%%%%%%%%%%%%%%%%
\MyTitle{Calculations} --- %%%
%%%%%%%%%%%%%%%%%%%%%%%%%%%%%%
%
LIS calculations of CR proton and helium are made using a spatial dependent \emph{two-halo model}
of CR propagation in the Galaxy \citep{Tomassetti2015TwoHalo,Feng2016,Evoli2017}
constrained against recent data from Voyager-1 and AMS.
From Voyager-1, we used all available data down to 100 MeV/n energies such as, in particular,
proton flux data at 140-320 MeV and helium flux data at 110-600 MeV/n \citep{Cummings2016}.
From AMS, we used data on primary CR spectra \p-\He-\C-\Oxy{} at $\R>$\,60\,GV \citep{Aguilar2015Proton,Aguilar2015Helium},
and measurements of the \BC{} ratio at $\R>$\,4\,GV \citep{Aguilar2016BC}. The minimal rigidities are chosen to
ensure that the measurements are unaffected by solar modulation \cite{Tomassetti2017BCUnc}.
In our model, the injection of primary CRs is parameterized by source terms $S\propto(\R/{\rm GV})^{-\nu}$
with $\nu=$\,2.28$\pm$0.12 for protons, and $\nu=$\,2.35$\pm$0.13 for $^{4}$\He{} and heavier primaries.
The transport in the $L$-sized Galactic halo is described by a diffusion coefficient $D = \beta D_{0}(\R/GV)^{\delta_{i/o}}$ 
with  $D_{0}/L=0.01\pm$0.002\,kpc/Myr.
The indices ${\delta_{i/o}}$ account for two different diffusion regimes: $\delta_{i}=0.18\pm$0.05 
in the near-disk region $|z|<\xi\,L$, and  $\delta_{o}=\delta_{i}+\Delta$
in the outer halo $|z|>\xi\,L$, where $\xi=0.12\pm$0.03 and $\Delta=0.55\pm$0.11.
Diffusive reacceleration is also considered.
The two-halo model of CR diffusion allows for only moderate values of Alfv{\'e}nic speeds, $v_{A}\cong$\,0--6\,km\,s$^{-1}$.
Other models, however, make use of stronger reacceleration \citep{Trotta2011,DruryStrong2017,EvoliYan2014}.
%%%%%%%%%%%%% p-He LIS fluxes %%%%%%%%%%%%%%%%%%
\begin{figure}[!t]
\includegraphics[width=0.44\textwidth]{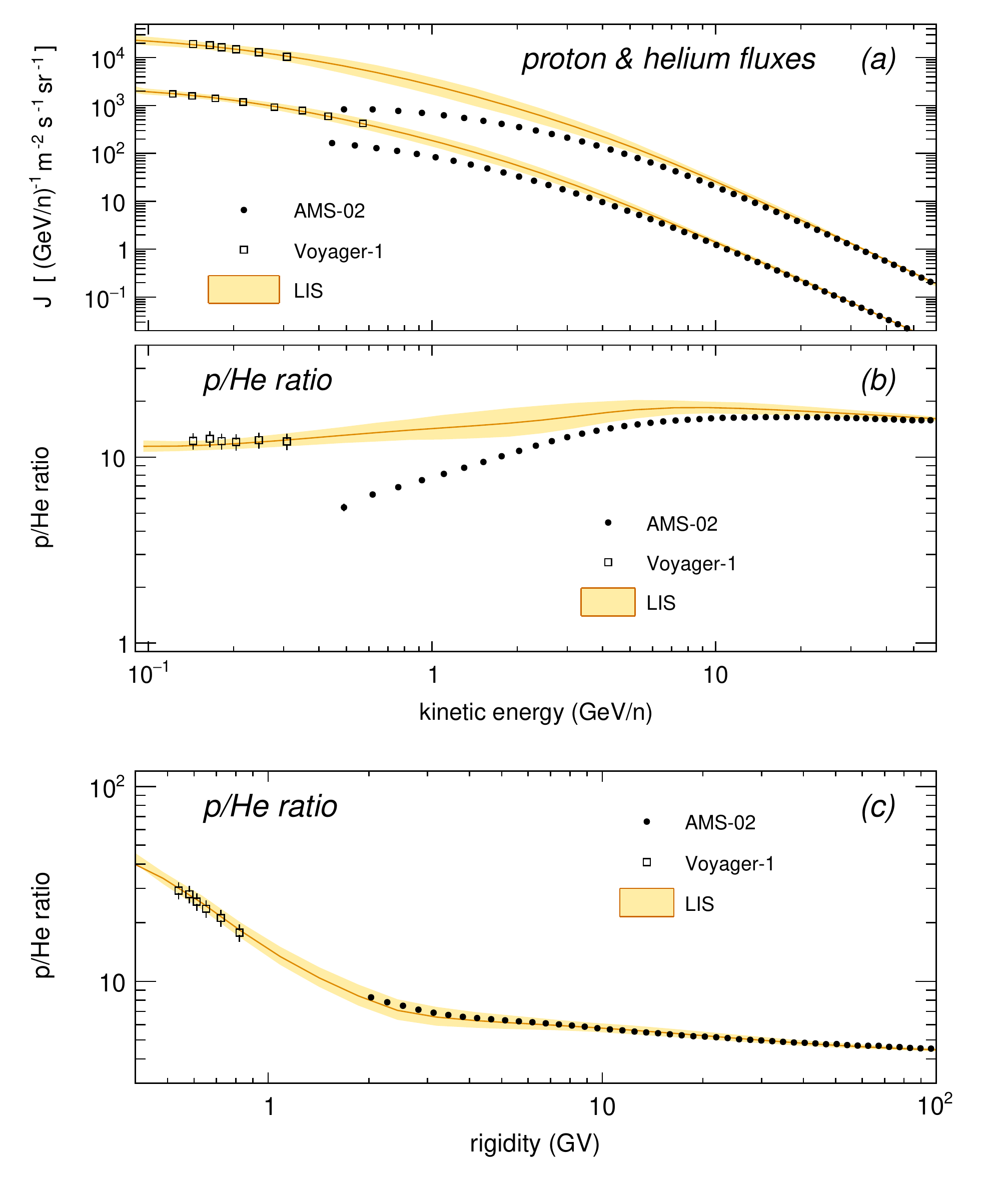}
\caption{ 
  (a) proton and helium LIS's in comparison with the data from AMS \citep{Aguilar2015Proton,Aguilar2015Helium} and Voyager-1 \citep{Cummings2016};
  (b) \pHe{} ratio as function of kinetic energy per nucleon;
  (c) \pHe{} ratio as function of rigidity.
}
\label{Fig::ccLISProtonHelium}
\end{figure}
%%%%%%%%%%%%%%%%%%%%%%%%%%%%%%%%%%%%%%%%%%%%%%%
% 
To model production and destruction of secondaries, improved evaluations of fragmentation cross sections
have been adopted, in particular for the \Be-\B{} isotopes \citep{Tomassetti2017BCUnc} 
(important for constraining the transport parameters with the \BC{} ratio),
and the $^{2}$\Hyd-$^{3}$\He{} isotopes \citep{Tomassetti2012Iso,TomassettiFeng2017}
(important for calculating LIS fluxes and uncertainties). 
Results are shown in Fig.\,\ref{Fig::ccLISProtonHelium}a, where the difference between
LIS calculations and AMS data shows effect of solar modulation and its energy dependence. 
The resulting 1-$\sigma$ uncertainties are shown as shaded bands. 
The largest uncertainties (up to $\sim$\,40\,\%) lie in the $\sim$\,1-10\,GeV energy region that is not covered by direct measurements.
This region is also sensitive to key parameters such as $\delta_{i/o}$, $D_{0}/L$, or $v_{A}$. 
It is important to note, however, that transport processes cause similar effects to protons and helium. Thus,
their uncertainties are tightly \emph{correlated} and partially canceled in the LIS \pHe{} ratio of Fig.\,\ref{Fig::ccLISProtonHelium}b,c. 
From Fig.\,\ref{Fig::ccLISProtonHelium}c, it is also shown that the \pHe{} ratio
measured inside the heliosphere is very similar to its interstellar value when it is represented as function of rigidity.
This explains why, in past measurements, no time variations were detected in the \pHe{} ratio \citep{Bindi2017}.

The subsequent transport of CRs in the heliosphere is described the Krymsky-Parker equation \citep{Krymsky1964,Potgieter2013,Moraal2013}.
In spherical symmetry, the equation reads:
\begin{equation}\label{Eq::TransportEquation}
\frac{\partial{\psi}}{\partial{t}}=\frac{1}{r^{2}}\frac{\partial}{\partial{r}}\left( r^{2}K\frac{\partial{\psi}}{\partial{r}}\right)
-V\frac{\partial{\psi}}{\partial{r}}
+\frac{1}{3r^{2}}\frac{\partial}{\partial{\R}} r^{2}V  \R\frac{\partial{\psi}}{\partial{\R}}
\end{equation}
where $\psi(t,r,\R)$ is the CR phase space density as function of time $t$, radial position ${r}$, and rigidity $\R$.
In the following, we set the wind speed at $V\cong$\,400\,km\,s$^{-1}$ and the modulation boundary at $d\cong$\,120\,AU.
At the boundary, the CR fluxes is imposed to match their LIS values of Fig.\,\ref{Fig::ccLISProtonHelium}.
For the diffusion, we adopt a benchmark form $K=K_{0}\beta \left(\R/{\rm{GV}}\right)$, with $K_{0}\equiv{10^{22}}k_{0}$\,cm$^{2}$s$^{-1}$,
where the changing modulation is captured by $k_{0}=k_{0}(t)$, a time-series of adimensional free parameters  \citep{Manuel2014,Bobik2016}.
Following earlier works \citep{Fisk1971}, we have computed numerical solutions for $\psi(r,\R)$ of all relevant
isotopes using the Crank-Nicolson method along a two-dimensional $r-\rm{ln}\R$ grid of 610$\times$500 nodes \citep{Thomas1995}.
The $k_{0}$ parameters and their uncertainties are constrained by means of the least-squares minimization method \citep{JamesRoos1975}.
We used the time-series of proton flux measured by AMS at $r=1$\,AU for 79 Bartel's
rotations \citep{Aguilar2018PHe}, ranging from $\R=1$ to $\R=60$\,GV, for a total of 3555 data points.
With this method, the temporal evolution of the proton flux near-Earth is obtained as a time series of
steady-state solutions of Eq.\,\ref{Eq::TransportEquation}.
The model is then used to compute the flux of other isotopes and, eventually, the \pHe{} ratio.

\begin{figure}[!t]
\includegraphics[width=0.44\textwidth]{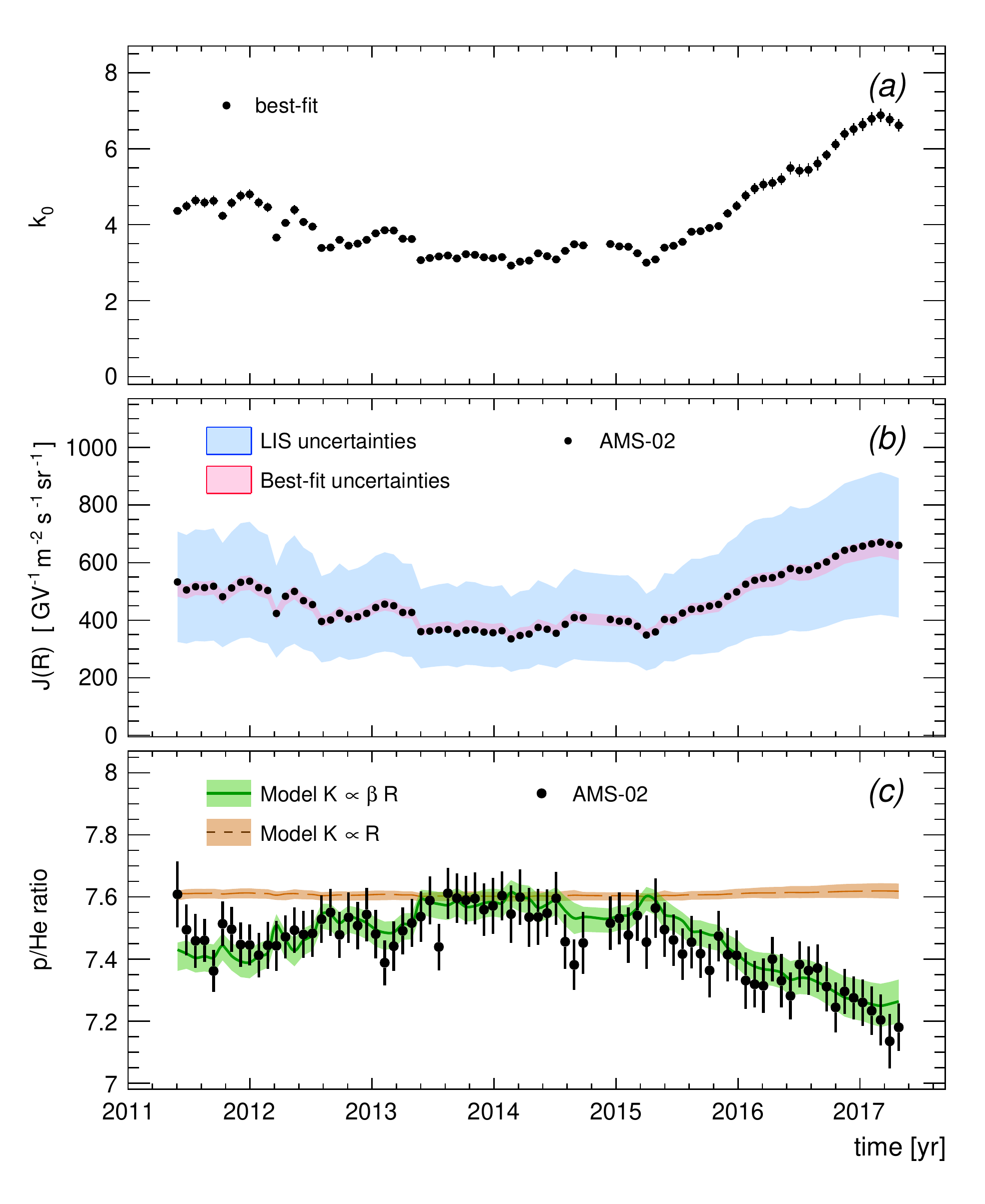}
\caption{ 
  (a) Time-series of the best-fit $k_{0}$-values derived with the AMS proton
  data at $\R=1-60$\,GV \citep{Aguilar2018PHe}; 
  (b) time profile of the proton flux measured by AMS 
  at $\R=$\,2\,GV in comparison with calculations, LIS uncertainties 
  and uncertainties from the fit;
  (c) measured time profile of the \pHe{} ratio at $\R=$\,2\,GV in comparison
  with best-fit calculations for $K=\beta\R$ (thick solid line)
  and $K\propto\R$ (thin dashed line).
}
\label{Fig::ccFitResultsVSTime}
\end{figure}

%%%%%%%%%%%%%%%%%%%%%%%%%%%%%%%%%%%%%%%%
\MyTitle{Results and discussion} --- %%%
%%%%%%%%%%%%%%%%%%%%%%%%%%%%%%%%%%%%%%%%
%
In Fig.\,\ref{Fig::ccFitResultsVSTime}, we show the  best-fit time-series of ${k}_{0}$ (a)
and the time profile of the CR proton fluxes at $\R=2$\,GV (b) in comparison with the AMS data.
%%%%%%%%%%%%% Fit results %%%%%%%%%%%%%%%%%%
%%%%%%%%%%%%%%%%%%%%%%%%%%%%%%%%%%%%%%%%%%%%
In the figure, the blue shaded band represents the uncertainty on the proton LIS directly 
propagated to the modulated spectra, once the ${k}_{0}$ parameters are fixed to their best-fit value.
It should be noted, however, that variations in the proton LIS are re-absorbed by the $k_{0}$ fitting. 
The relevant uncertainties are shown by the pink shaded band.
The comparison of the two bands illustrates the \emph{potential} level of precision
to which our LIS knowledge could be reached after a proper account of the modulation effect. 
%%%%%%%%%%%%% p/He ratio VS time %%%%%%%%%%%%%%%%%%
\begin{figure*}[!t]
\includegraphics[width=0.90\textwidth]{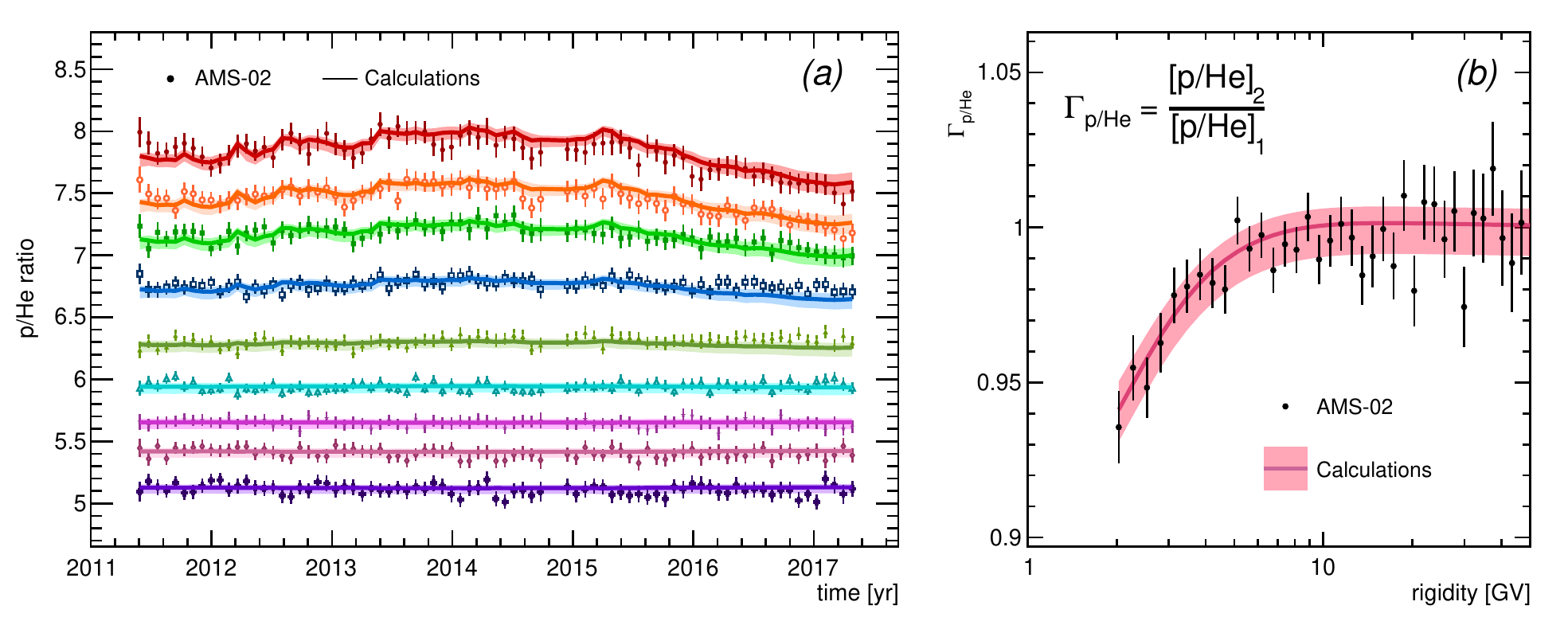} 
\caption{ 
  (a) time profiles of the \pHe{} ratio evaluated at rigidities $\R=$\,2.2, 2.5, 2.8, 3.4, 5.1, 7.4, 10.5, 15, and 22 GV (from top to bottom); %14.7
  (b) rigidity dependence of the ratio $\Gamma_{\rm p/He}= {[\pHe]_{t_{2}}}/{[\pHe]_{t_{1}}}$
  calculated for February 2014 ($t_{1}$) and May 2017 ($t_{2}$).
  In both plots, the new data from AMS are compared with model predictions and their uncertainties.
}
\label{Fig::ccPHeRatioVSTime}
\end{figure*}
%%%%%%%%%%%%%%%%%%%%%%%%%%%%%%%%%%%%%%%%%%%%%%%%%%%
Our prediction for the temporal dependence of the \pHe{} ratio at $\R\approx$\,2\,GV is
shown in Fig.\,\ref{Fig::ccFitResultsVSTime}c, where the total \He{} flux is
computed using the proton-driven constraints. This illustrates the main result of this \Paper.
The uncertainty bands account for the different time-dependence arising from the LIS models at varying propagation parameters, 
the uncertainties on their isotopic composition, and the errors associated with the $k_{0}$ fitting.
It can be seen that the long-term time behavior of the \pHe{} ratio is predicted very well.
We also note that the long-term behavior of the \pHe{} ratio is correlated with the 
absolute intensity of the two fluxes which, in turns, reflects the level of solar activity. 
Appreciable variations in the \pHe{} ratio are observed after May 2015 during the so-called
\emph{recovery phase} when the primary fluxes increase rapidly by nearly a factor of $2$.
We also note that this phase starts about one year after the maximum
of Solar Cycle 24, reflecting a time lag in CR modulation \citep{Tomassetti2017TimeLag}.
For sake of comparison, we also report calculations arising from a diffusion coefficient of the type $K{\propto}k_{0}\R$,
\ie, without the $\beta$-factor (thin dashed line). 
In this case, the predicted \pHe{} ratio is remarkably constant showing that the difference in the
proton and helium LIS's plays a minor effect (and opposite to the observed \pHe{} trend \citep{Gieseler2017}).
Our results are summarized in Fig.\,\ref{Fig::ccPHeRatioVSTime}a where the \pHe{} time profile is shown for several rigidity values.
To show how the \pHe{} structure depends upon rigidity, in Fig.\,\ref{Fig::ccPHeRatioVSTime}b we compute
the observable $\Gamma_{\rm p/He}\equiv{[\pHe]_{t_{2}}}/{[\pHe]_{t_{1}}}$ for the reference dates February 2014
($t_{1}$, where the \pHe{} ratio is maximum) and May 2017 ($t_{2}$, where the ratio is minimum).
Deviations from a constant $\Gamma_{\rm p/He}$ value are predicted to appear below a few GV, in full agreement with the data.

Qualitatively, these results can be understood under the simple framework
of the \emph{convection-diffusion approximation} (CDA) \citep{Moraal2013}.
For $j$-type CRs entering the heliosphere with a LIS $J_{j}^{0}(\R)$, their modulated flux is approximately given by
$\sim J_{j}^{0}e^{-\mathcal{M}}$, where the adimensional parameter $\mathcal{M}=\int_{r_{0}}^{d}\frac{V}{K}dr\propto\frac{Vd}{K_{0}}$ sets the level of modulation.
This shows clearly the parameter degeneracy between $K$ and $V$.
An important quantity for the phenomenology is the combination $\mu{\equiv}Vd/K_{0}$,
which captures the effects of changing conditions in solar activity.
For $\mathcal{M}\lesssim$\,1, the \pHe{} ratio is expected to vary as:
\begin{equation}\label{Eq::CD}
\pHe \approx \frac{J_{\rm p}^{0}}{J_{\rm He}^{0}} \left\{ 1 -\frac{\mu(t)}{\R/{\rm GV}}\left[\frac{1}{\beta_{\rm p}(\R)}-\frac{1}{\beta_{\rm He}(\R)} \right] \right\}\,,
\end{equation}
where $\beta(\R)= \R / \sqrt{ \R^{2} + (m_{p}A/Z)^{2}}$.
Similarly, the $\Gamma_{\pHe}(\R)$ function can be readily calculated:
\begin{equation}
\Gamma_{\pHe}(\R) \approx
\frac{
1-\frac{\mu(t_{2})}{\R/{\rm GV}}\left[\frac{1}{\beta_{\rm p}(\R)}-\frac{1}{\beta_{\rm He}(\R)}\right]
}{
1-\frac{\mu(t_{1})}{\R/{\rm GV}}\left[\frac{1}{\beta_{\rm p}(\R)}-\frac{1}{\beta_{\rm He}(\R)}\right]
}
\end{equation}
These simple relations explain why the modulated \pHe{} ratio increases with increasing level
of modulation and vice-versa, and why the effect is more pronounced at nonrelativistic rigidities.
In the relativistic limit $\beta_{\rm p}\approx\beta_{\rm He}\approx$\,1, one
recovers $\pHe \approx J_{\rm p}^{0}/J_{\rm He}^{0}$ and $\Gamma_{\pHe}\approx$\,1,
so that the modulated ratio becomes representative of its interstellar value, as also 
suggested by Fig.\,\ref{Fig::ccLISProtonHelium} and Fig.\,\ref{Fig::ccPHeRatioVSTime}. 
It is also important to note that the specific rigidity dependence of the CR mean free path does not alter 
the basic predictions, as long as $\lambda(\R)$ is identical for protons and \He.
Similar considerations were made in past works to argue that CRs in the heliosphere
follow a $K\propto\beta\R$ diffusion, as distinct from $K\propto\R$ \citep{Jokipii1967,Gloeckler1967,Biswas1967}.
From Eq.\,\ref{Eq::CD}, one may note that the  p/He interstellar ratio is factorized
out, and similarly, the  $\Gamma$-function is LIS-independent.
This is a consequence of the CDA neglection of energy changes.
We also stress, however, that the $\mu(t)$ time-series \emph{does} depend
on the assumed LIS if the modulation parameters are determined from the data.
In our calculations, however, we opted for the numerical method, because the CDA solution
is unsuitable for the precision demanded by the AMS data \citep{Moraal2013}.
Using the recent observations and an improved Galactic transport model,
we followed a multi-channel and data-driven approach, in which we have adopted the minimal
set of parameters that capture the phenomenology of the \p-\He{} modulation.
In the interest of simplicity, we have neglected the tensor nature of CR diffusion (that would require a more complex modeling)
and we did not attempt to relate $K(\R)$ with the properties of the heliospheric plasma (that may require an explicit time-dependent
description of the problem \citep{Strauss2011}). Hence, the diffusion coefficient obtained in this work should be regarded
as an effective quantity, representing the space- and time-averaged CR propagation histories at a given reference epoch.
To test the robustness of our findings, we have repeated our calculations using several LIS's proposed recently
\citep{Tomassetti2017TimeLag,Tomassetti2017BCUnc,Tomassetti2015PHeAnomaly,Corti2016},
all giving consistent results within the uncertainties. We have also tested generalized mean free paths, including power-law or double power-law behavior,
radial dependence, or drift terms  \citep{Moraal2013,Potgieter2013,Boschini2017,Jokipii1977,IsenbergJokipii1978}. 
While the use of complex descriptions (and more parameters) may improve the global fits in the CR fluxes,
the predicted evolution of the \pHe{} is found to be insensitive to the exact $\lambda(\R)$ dependence
as long as it is a unique function of rigidity.
In summary, the key feature to explain the long-term behavior of the \pHe{} ratio is a factorized form of
CR diffusion, $K\propto\beta(\R)\times\lambda(\R)$, that gives rise to a $Z/A$-dependent modulation effect.
\\[0.15cm]
%%%%%%%%%%%%%%%%%%%%%%%%%%%%%
\MyTitle{Conclusions} --- %%%
%%%%%%%%%%%%%%%%%%%%%%%%%%%%%
%
This work is aimed at interpreting the long-term behavior of the \pHe{} ratio recently observed by AMS.
To describe the data at precision level demanded by AMS, we took advantage of recent developments in CR observations and modeling.
The \p-\He{} LIS's, their isotopic composition, and their uncertainties are calculated using improved models of
CR propagation in Galaxy constrained against new Voyager-1 data \citep{Cummings2016},
and high-energy measurements from AMS \citep{Aguilar2015Proton,Aguilar2015Helium}.
The time-dependent effects of solar modulation are described by numerical calculations
of CR transport in the heliosphere calibrated against the new time-resolved data from AMS on CR protons \citep{Aguilar2018PHe}.
Our data-driven approach enabled us to predict the temporal dependence of the \pHe{} ratio and to test specific forms of the diffusion coefficient.
We have shown that the long-term behavior of the \pHe{} ratio arises naturally from the $\beta\times\lambda(\R)$ dependence of CR diffusion in the heliosphere.
These findings support the concept of \emph{universality} in the CR propagation histories in the heliosphere,
expressed by their mean free path $\lambda(\R)$ that is a unique rigidity-dependent function for all charged nuclei.
Here the test was performed for the most extreme case of the \pHe{} ratio.
Further tests can be made with other species such as $^{2}$\Hyd{}, $^{3,4}$\He, Li-Be-B or C-N-O, that are being measured by the AMS experiment. 
\\[0.15cm]
%
%%%%%%%%%%%%%%%%%%%%%%%%%%%%
\MyTitle{Note added} --- %%%
%%%%%%%%%%%%%%%%%%%%%%%%%%%%
%
While this work was in review, we became aware of a related study from \citet{Corti2018}.
Their work is based on the same data sets of ours, but it follows different approaches for the Galactic and heliospheric transport modeling.
Their results are consistent with those presented in this Letter.
\\[0.2cm]
\footnotesize{%
  We thank our colleagues of the AMS Collaboration for valuable discussions.
  The data analyzed in this work are publicly available at the \textsf{SSDC Cosmic-Ray Database}
  hosted by the \emph{Space Science Data Center} of the \emph{Italian Space Agency}
  \citep{DiFelice2017}.
  N.T. and B.B. acknowledge the European Commission for support under the H2020-MSCA-IF-2015 action, grant No.\,707543
  MAtISSE -- \emph{Multichannel investigation of solar modulation effects in Galactic cosmic rays}.
}

%%%%%%%%%%%%%%%%%%%%%%%%%%%%%%%
 %%%
%%%%%%%%%%%%%%%%%%%%%%%%%

%%%%%%%%%%%%%%%%%%%%%%%%%%%%%%%%%%%%%%%%%%%%%%%%%%%%%%%
\end{document}